\documentclass[aip,superscriptaddress,nofootinbib,showpacs,amssymb,graphicx]{revtex4-1}
\usepackage{graphicx}
\usepackage{epstopdf}
\input epsf
\usepackage{epsfig}
\usepackage{caption}
\usepackage{bm}
\usepackage{amsmath}
\usepackage{amssymb}

\let\xxxhat\hat
\renewcommand{\hat}[1]{{\mbox{\boldmath $ {\xxxhat {#1}} $}}}

\begin{document}
\title{Reconstruction of 3D Image of Nanorice Particle from Randomly Oriented Single-Shot Experimental Diffraction Patterns Using Angular Correlation Method}
\author{S. S. Kim}
\affiliation{Department of Physics, University of Wisconsin-Milwaukee
P. O. Box 413, Milwaukee, WI 53201, USA}
\author{P. Nepal}
\affiliation{Department of Physics, University of Wisconsin-Milwaukee
P. O. Box 413, Milwaukee, WI 53201, USA}
\author{D. K. Saldin}
\affiliation{Department of Physics, University of Wisconsin-Milwaukee
P. O. Box 413, Milwaukee, WI 53201, USA}
\author{C. H. Yoon}
\affiliation{Linac Coherent Light Source, SLAC National Accelerator Laboratory, Menlo Park, CA 94025, USA}

\begin{abstract}

We reconstructed intensities in Fourier space and electron densities in the real space for an azimuthally symmetric object Nanorice particle (Iron Oxide nanoparticle) exposed in the ultrashort, bright and coherent x-ray free electron laser (XFEL) pulses with random unknown orientations through the method of angular correlations among intensities appeared in ninety eight 2D diffraction patterns collected at Linac Coherent Light Source (LCLS).

\end{abstract}

\maketitle
An X-ray Free Electron Laser (XFEL) is now generating X-ray pulse trains of unprecedented brilliance of about 10 billion times of what was previously possible with the rate of only a few femtosecond scale [1]. As such it has given rise to the speculation that it may be possible to determine the structures of uncrystallized individual biomolecules such as proteins and viruses. Although the ultimate aim is to determine the structures of biomolecules, it would be directive for the 3D reconstruction work to demonstrate the feasibility of the approach to simpler objects initially. There has been some work already on reconstructing prolate spheroids [2] of metallic particles by reconstructing 3D Fourier intensities of a large and simple Iron Oxide nanoparticle ($Fe_2O_3$ coated with $SiO_2$ : Nanorice particle).

In this paper, we examine the capabilities of the angular correlation method that is based on the angular momentum decomposition of scattered intensities, which enables us to overcome common problems such as missing or imperfect data, effect of noise, curved Ewald sphere, shot to shot incident X-ray pulse intensity variations that are inevitable in experiments. The method of angular correlation recovers quantities from single-shot experimental diffraction patterns (DPs) of randomly oriented particles, as expected to be measured at an XFEL proportional to quadratic functions of the spherical harmonic expansion coefficients of 3D intensity distribution (so called diffraction volume) of a single particle. This method consists first of the reconstruction of the diffraction volume. Conventionally, this is followed by the reconstruction of a real space image by an iterative phasing algorithm [3][4]. We have previously shown that it is possible to reconstruct 3D images of a randomly oriented icosahedral or helical virus from the average over all measured diffraction patterns of such correlations through simulation diffraction patterns using the samples in the protein data bank [5][6].

Although it is true that some methods have reconstructed the particle to the resolution available to the experiment, we indicate the advantage of the angular correlation method. This is a piece of information given to the algorithm in our case, namely the azimuthal symmetry of the Nanorice particle which is known beforehand. This extra information (often in an angular momentum basis) could make the difference in 3D conformational reconstruction image of an object between the work with a large number of diffraction patterns (at least a couple of thousand or many more) and a small number of those (a few hundreds). It is also conspicuous that such widely different approaches give rise to essentially the same results, lending more credibility to the structures recovered. In addition, currently no other method can work appropriately for making 3D reconstruction image in the real space of a nano particle whose size is about 200nm with less than a hundred experimental DPs as we achieved here. It should be stressed that the method described in this paper is flexible enough to reconstruct the structure from  ``single-particle" experiments such as in the recent Single Particle Initiative (SPI) at the LCLS. What is used here is the angular correlations, which are the averages over the products of two intensities lying on an equidistant polar grid ring along monotonously increasing angles between two intensities collected from each DP.

\begin{figure}
\centering
\includegraphics[scale=0.5]{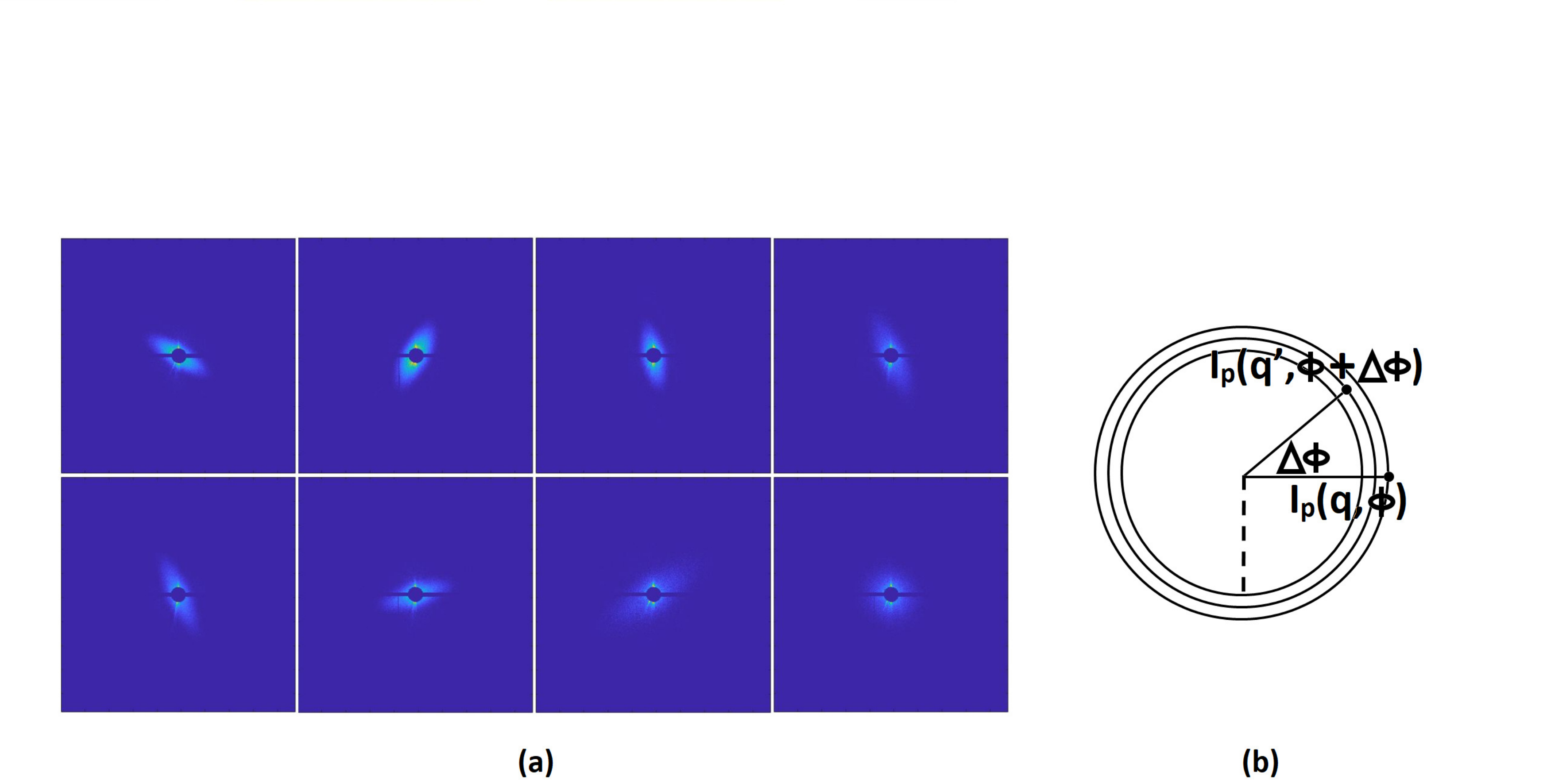}
\label{FIG.1}
\caption{(a) 8 of 98 randomly oriented Single-Shot 2D experimental diffraction patterns of a Nanorice particle (Iron Oxide nanoparticle). (b) Two intensities on the polar coordinate separated by an angular distance $\Delta\phi$. The average of these products forms an angular correlation $C_2(q,q',\Delta\phi)$ and two-point angular triple correlations $C_3(q,q',\Delta\phi)$.}
\end{figure} 

We performed our experiment at LCLS facility with each XFEL pulse whose photon energy is 1.2KeV, a unit cell size of $75 \mu$m in the detector and the distance between the sample and the detector is 75cm. FIG.2 shows a brief diagram of experimental setup to collect diffraction patterns through the detector at LCLS. In our work for reconstructing 3D image of 
\begin{figure}
\centering
\includegraphics[scale=0.5]{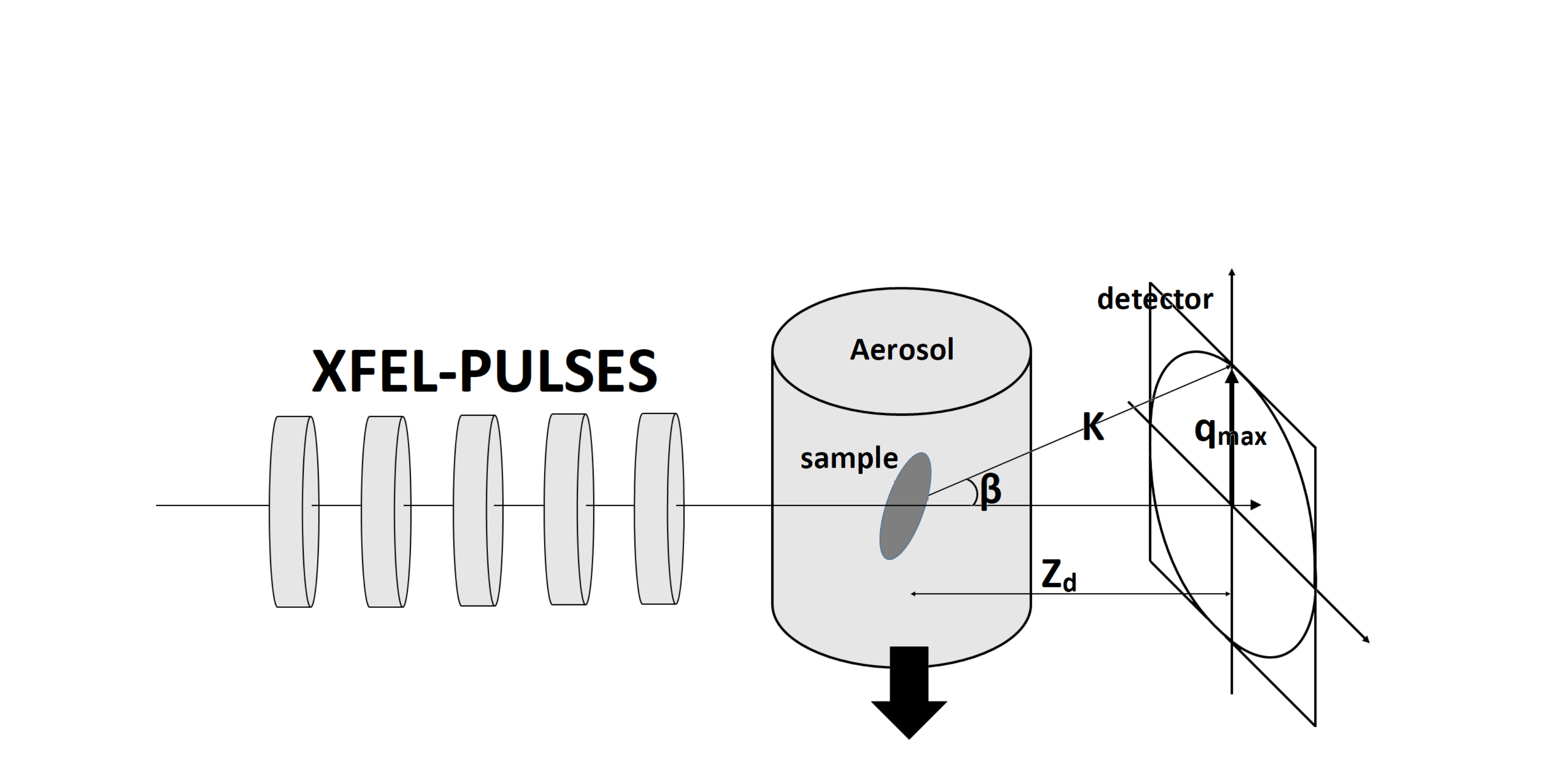}
\label{FIG.2}
\caption{Brief diagram of experimental setup to collect diffraction patterns of Nanorice particle through 1039x1024 pixels in the detector at LCLS. $Zd$(distance between the sample and the detector)=$75$cm, $\Delta p$(pixel size)=7.5x$10^{-5}$m, $K=2\pi/ \lambda=104.8nm^{-1}$}
\end{figure} 
the Nanorice particle, we selected the single shot diffraction patterns from multiple shot ones since the DPs provided by LCLS have already been sorted [7]. To apply the angular correlation theory to the experimental data, we have to find an important parameter $q_{max}$ in our procedure via the experimental setup in FIG.2 by

\begin{equation}
q_{max}=Ksin\beta=\frac{2\pi}{\lambda}sin \left[ tan^{-1} \left(\frac{N\Delta p}{2Zd} \right) \right]
\end{equation}
where $N$=the number of total pixels on a vertical line of the detector=1024, Zd=distance between the sample and the detector=75 cm and $\Delta p$=pixel size=7.5x$10^{-5}m$.\\

\begin{figure}
\includegraphics[scale=0.5]{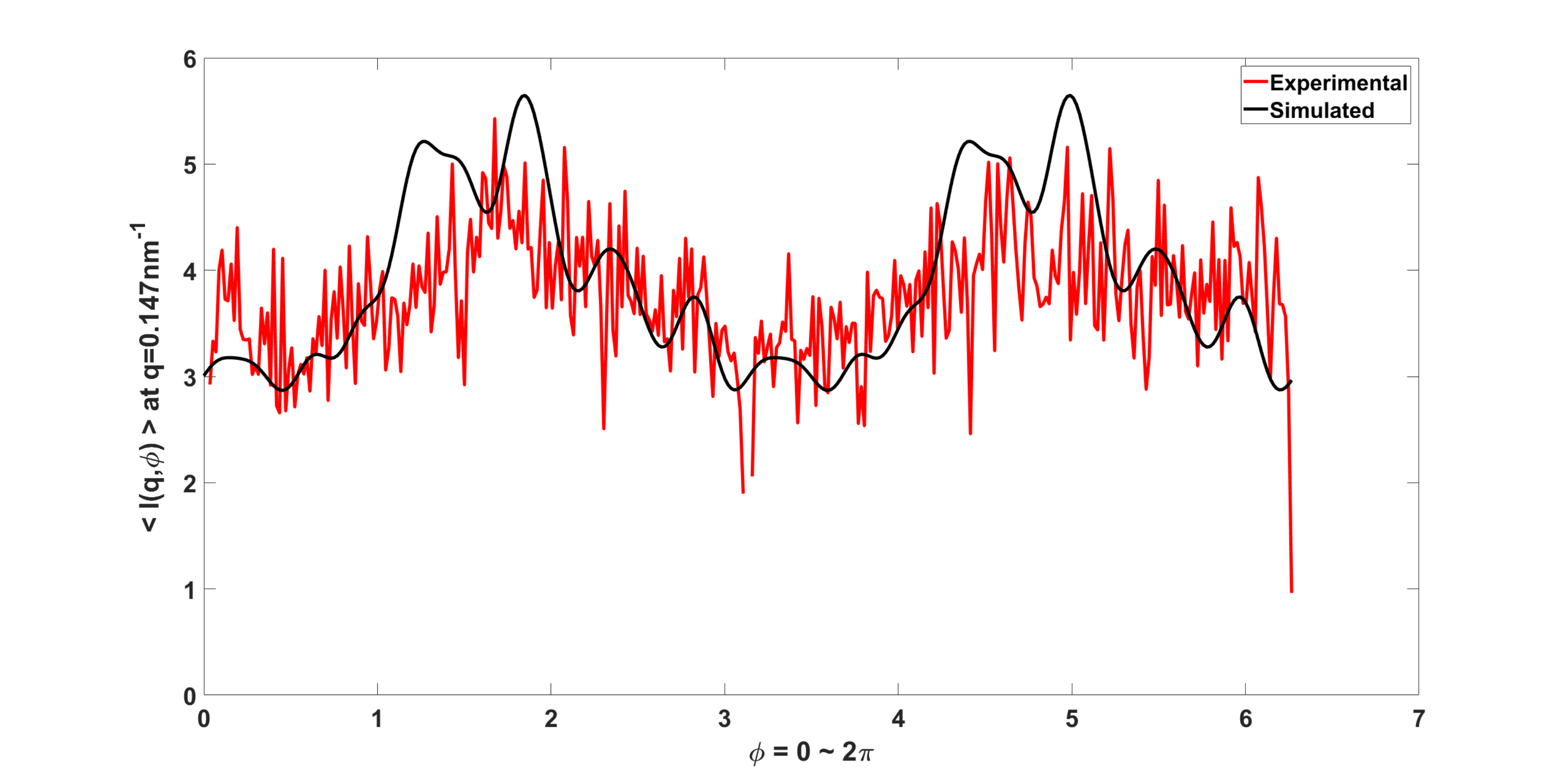}
\label{FIG.3}
\caption{The plot of  $<I(q,\phi)>_{\phi}$ vs. $\phi$. Along the angular positions from $\phi=0$ to $\phi=2\pi$ where $q=0.147nm^{-1}$ away from the center. The simulated data are scaled for comparison. It is observed that the average intensities of each angular position along $q=0.147nm^{-1}$ circles of 98 DPs, experimental data fluctuate unlike simulated ones. This causes the peaks at $\Delta\phi=0$ in $C_2(q,q,\Delta \phi)$ and $C_3(q,q,\Delta \phi)$ graphs.}
\end{figure}

\indent The first step in using this method is to calculate angular cross correlations on each DP as in FIG 1(a) in polar coordinates as in FIG 1(b). Polar coordinates are natural for this 3D imaging process since the particles differ mainly in their orientations. They may also differ in their positions, but this does not affect the DP intensities that are insensitive to the phases of the scattered amplitudes. This is relevant so long as the particle is in the pulse at a time. Otherwise the intensities are sensitive to the relative displacements of the particles in the same DP. Even in this case one might hope that due to the random nature of these displacements, such relative phases are unimportant [8]. The angular pair correlations are defined by

\begin{equation}
C_2(q,q',\Delta \phi)=\displaystyle{\frac{1}{N}\sum\limits_{p=1}^{n} \sum\limits_{\Delta\phi=0}^{2\pi}I_p(q,\phi)I_p(q',\phi+\Delta\phi)}
\end{equation}

where $I_p(q,\phi)$ is the measured intensity at a resolution ring $q$ and azimuthal angle $\phi$ on a diffraction pattern $p$, \textit{n} is the number of DPs, and $I_p(q',\phi+\Delta\phi)$ the corresponding intensity at a resolution ring $q'$, azimuthal angle $\phi+\Delta\phi$, $N$ is the total number of two intensity products. It is noted that in the middle of the detector from the real experiment, unlike simulated data zero intensities appear everywhere  that make the products of two intensities zeros. This may cause the value of the average of the intensity product $C2$ and $C3$ inaccurate. Thus we only let the nonzero intensity products participate in calculation for the preciseness of  $C2$ and $C3$ . We chose $\Delta\phi$ is an angle increased by $1^o=\pi/180$ rad from 0 to 2$\pi$. Similar to the pair correlations, two-point angular triple correlations may be defined by

\begin{equation}
C_3(q,q',\Delta \phi)=\displaystyle{\frac{1}{N}\sum\limits_{p=1}^{n} \sum\limits_{\Delta\phi=0}^{2\pi}I_p^2(q,\phi)I_p(q',\phi+\Delta\phi)}
\end{equation}

Through our work, we use only diagonal parts where $q=q'$ to form $C_2(q,q,\Delta\phi)$ and $C_3(q,q,\Delta\phi)$.
In experimental data, unlike simulated ones, there are peaks at $\Delta\phi=0$ in $C_2(q,q,\Delta\phi=0)$ and $C_3(q,q,\Delta\phi=0)$ since the intensities along a resolution ring $q$ from experimental data fluctuate saw-like as in FIG 3, and consequently generates the $B_l(q,q)$ peaks at $l=0$ different from the simulated ones. Thus we removed the peaks from $C2$ and $C3$. Namely, $C_2(q,q,\Delta\phi=0^{\circ})=C_2(q,q,\Delta\phi=1^{\circ})$ and $C_3(q,q,\Delta\phi=0^{\circ})=C_3(q,q,\Delta\phi=1^{\circ})$ as in FIG 4. 

\begin{figure}
\includegraphics[scale=0.5]{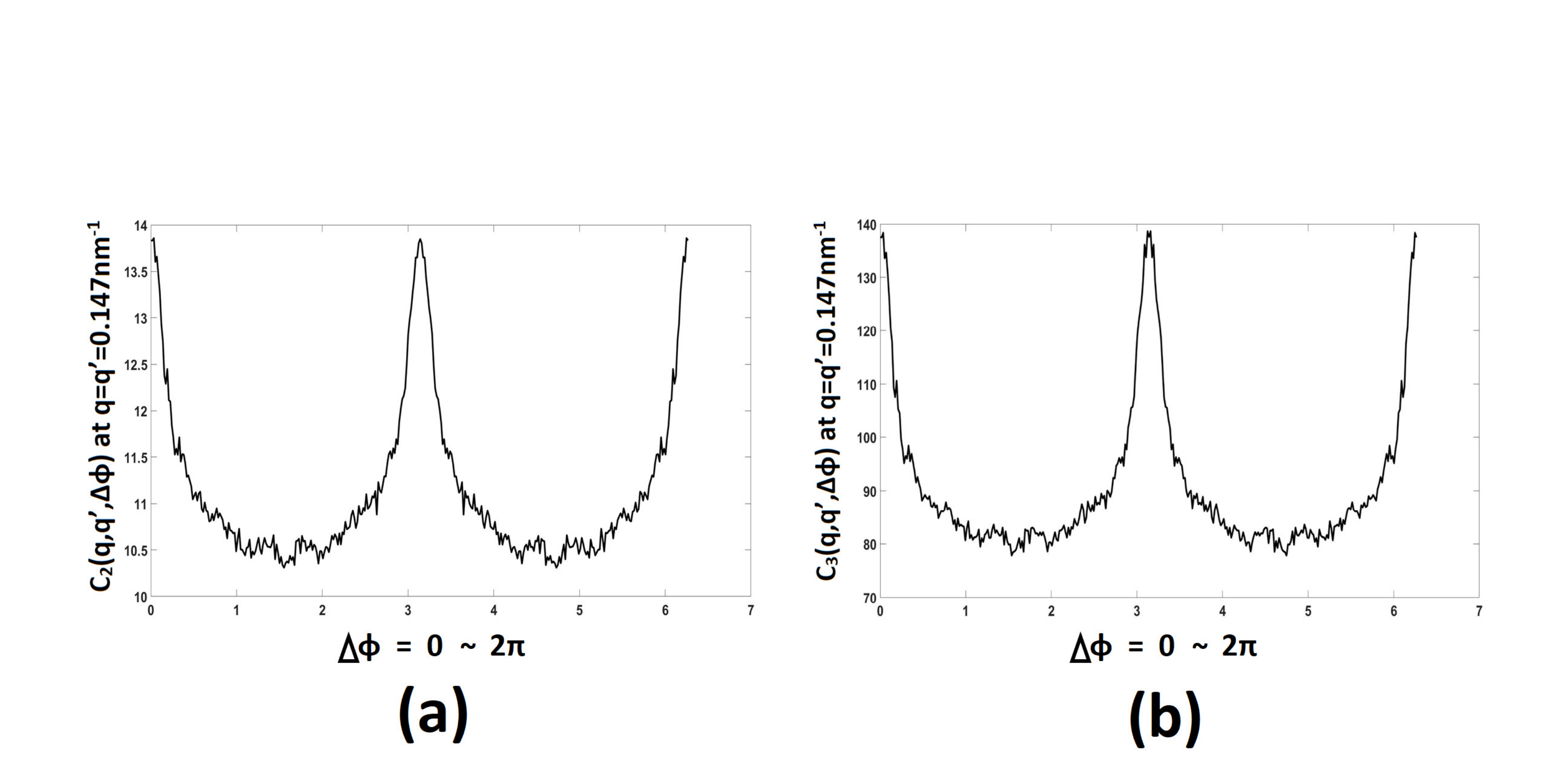}
\label{FIG.4}
\caption{(a) The plot of $C_2(q,q',\Delta \phi)$  vs. $\Delta\phi=0$ to $2\pi$. (b)  $C_3(q,q',\Delta\phi)$  vs. $\Delta\phi=0$ to $2\pi$. Both are at $q=q'=0.147nm^{-1}$. Both peaks from (a) and (b) at $\Delta\phi=0$ are removed.}
\end{figure}

The next step is to calculate other quantities necessary to generate the diffraction volume using these $C2$ and $C3$ in the following way [9],
\begin{equation}
B_l(q,q')=\displaystyle{\frac{2l+1}{2}\sum\limits_{\Delta\phi=0}^{\pi}P_l(cos\Delta\phi)sin(\Delta\phi)  C_2(\Delta\phi)d(\Delta\phi)}
\end{equation}

and

\begin{equation}
T_l(q,q')=\displaystyle{\frac{2l+1}{2} \sum\limits_{\Delta\phi=0}^{\pi}P_l(cos\Delta\phi)sin(\Delta\phi)  C_3(\Delta\phi)d(\Delta\phi)}
\end{equation} \\

where $C_2(\Delta\phi)=C_2(q,q',\Delta\phi)$, $C_3(\Delta\phi)=C_3(q,q',\Delta\phi)$, and we chose $d(\Delta\phi)=\pi/180$. $P_l$ is the Legendre polynomials. 

\begin{figure}
\includegraphics[scale=0.2]{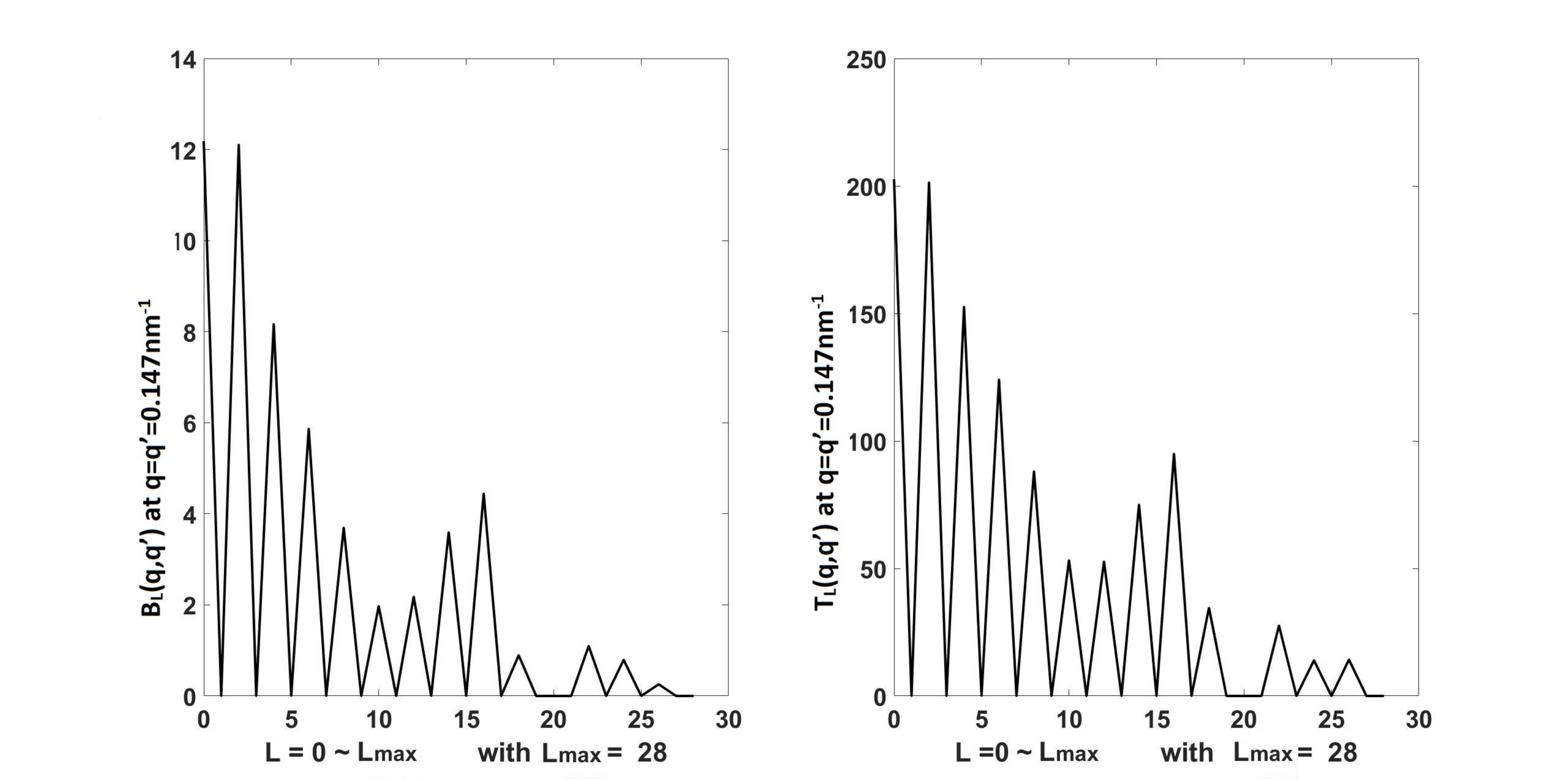}
\label{FIG.5}
\caption{(a) The plot of $B_l(q,q)$  vs. $\textit{l}$. (b) $T_l(q,q)$  vs. $\textit{l}$. at  $q=0.147nm^{-1}$ with $l=0,1,2, ... , 28$. Both peaks in $B_{l=0}(q,q)$ and $T_{l=0}(q,q)$ are adjusted by the ratio of $B_{l=0}/B_{l=2}$ and $T_{l=0}/T_{l=2}$ from simulated data.}
\end{figure} 

The angular pair correlations can be related to their angular momentum decomposition $B_l$ [10] by 

\begin{equation}
C_2(q,q',\Delta \phi)= \sum\limits_{l=0}^{l_{max}} F_l(q,q',\Delta\phi)B_l(q,q')
\end{equation}

where 

\begin{equation}
F_l(q,q',\Delta\phi)=\frac{1}{4\pi}P_l[cos\theta(q)cos\theta(q')+sin\theta(q)sin\theta(q')cos(\Delta\phi)]
\end{equation}

On the assumption that the Ewald Spheres are flat [10] as a 2D detector in the experiment,

$\theta(q)=\displaystyle{\frac{\pi}{2}-sin^{-1}}\left[\frac{q}{2k}\right]\approx\frac{\pi}{2}$ 

Thus,

\begin{equation}
 F_l(q,q',\Delta\phi)=\frac{1}{4\pi}P_l[cos(\Delta\phi)] 
\end{equation}

By the same way, the triple correlations defined by (3) can be written as

\begin{equation}
C_3(q,q',\Delta \phi)=\sum\limits_{l=0}^{l_{max}}F_l(q,q',\Delta\phi)T_l(q,q')
\end{equation}

with

\begin{equation}
T_l(q,q')=\sum\limits_{\substack{l_1,l_2 \\ m_1,m_2 \\ m}}^{l_{max}}G(l_1m_1,l_2m_2,lm)I_{l_1m_1}(q)I_{l_2m_2}(q)I_{lm}^{*}(q') 
\end{equation}

where $G$ is a Gaunt coefficient [11] and

\begin{equation}
I_{lm}(q)=\sum\limits_{\theta=0}^{\pi}\sum\limits_{\phi=0}^{2\pi}I(q,\theta,\phi)Y_{lm}^{*}(\theta,\phi)sin\theta d\theta d\phi
\end{equation}

The coefficient of $I_{lm}$ of a spherical harmonic expansion for the diffraction volume depend on the orientation of the diffraction volume relative to chosen z-axis. By choosing z-axis at the center of azimuthal symmetry, we eliminate the other components of $I_{lm}$ except $m=0$. We choose the orientation with the major axis of the ellipsoid along the z-axis. We accept this orientation by assuming that only $m=0$ components of $I_{l,m=0}(q)$ exist. Considering the distance between the sample and the detector, we can choose $q_{max}\approx0.3nm^{-1}$ using (1). Also one can choose the non degenerate $l$-values up to 28 in spherical harmonic coefficients as $l_{max}=28$. If $q_{max}$ is the maximum value of the reciprocal space coordinates q up to which the reconstruction is valid, conventional wisdom suggests that $l_{max}$ and $q_{max}$ should be related by $q_{max}R=l_{max}$ [12] where $R$ is the radius of the particle and the Nanorice particle has $R\approx 100nm$.
 
At this point these coefficients depend only on $l$ since we choose for all $m=0$ for the azimuthal symmetry. The magnitudes of these spherical harmonic coefficients are determined from [6]

\begin{equation}
I_{l,m=0}(q)=\sqrt{B_l(q,q)}
\end{equation}

Saw-like behavior of $C_2$ and $C_3$ in FIG 3. affects the peaks at $B_l$ and $T_l$ at $l=0$. Thus we adjust the peaks by adopting the ratio $B_{l=0}/B_{l=2}$ and $T_{l=0}/T_{l=2}$ from simulated data into experimental $B_{l=0}$ and $T_{l=0}$ as in FIG.5. All the above discretized expressions that help reproducing the same results we have here can be conceptually described as continuous forms [13].  From (12), the only unknown is the signs of $I_{l0}(q)$. Since $\sqrt{B_l(q,q)}$ provides both $\pm$ signs, using the values of $T_l(q,q)$ of the triple correlations calculated directly from the diffraction patterns of random particle orientations, the signs of $\sqrt{B_l(q,q)}$ can be determined by sequentially exhaustive searching the closest $T_{l0}$ values from between (10) and (5). After obtaining the signs of $I_{l0}(q)$, the diffraction volume $I(\mathbf{q})$ can be calculated from 

\begin{equation}
I(\mathbf{q})=\sum\limits_{l=0}^{l_{max}=28}I_{l0}(q)Y_{l0}(\theta,\phi)
\end{equation}

An iterative phasing algorithm [3][4] and a constraint for the azimuthal symmetry $m=0$ once in (14) in the real space is applied to this diffraction volume can then recover the electron density of the particle. 

\begin{equation}
\rho_{l,m=0}(r)=\sum\limits_{\theta=0}^{\pi}\sum\limits_{\phi=0}^{2\pi}\rho(r,\theta,\phi) Y_{lm=0}^{*}(\theta,\phi)sin\theta d\theta d\phi
\end{equation}

The diffraction volume of the particle is displayed in FIG.6(a) and the reconstructed electron densities are shown after phasing with the constraint as in FIG.6(b). This reconstruction image formed by (15) appears after phasing by removing nonazimuthal parts from the electron densities $\rho(r,\theta,\phi)$ using (14). 

\begin{equation}
\rho(\mathbf{r})=\sum\limits_{l=0}^{28}\rho_{l,m=0}(r) Y_{lm=0}(\theta,\phi)
\end{equation}

If we consider a generally used expression (16) for amplitudes in crystallography 

\begin{equation}
A(\mathbf{q})=\displaystyle{\int \rho(\mathbf{r})e^{-i\mathbf{q}\cdot\mathbf{r}}d^3\mathbf{r}=\sum\limits_{l,m}A_{l,m}(q)Y_{l,m}(\theta,\phi)}
\end{equation}

and take the only azimuthal parts by using $m=0$, then the azimuthal parts of the amplitude would be described as (17) through (15),

\begin{equation}
A_{l,m=0}(q)=\displaystyle{4\pi i^l\int\limits_{r=0}^{\infty}\rho_{l,0}(r)j_l(|qr|)r^2dr}
\end{equation}

where $j_l(|qr|)$ is a spherical Bessel function of order $l$ [14]. This $A_{l0}(q)$ components give rise to $A(\mathbf{q})$ by the way of (18).

\begin{equation}
A(\mathbf{q})=\sum\limits_{l,m=0}^{l_{max}=28}A_{l0}(q)Y_{l0}(\theta,\phi)
\end{equation}

Once we found $A(\mathbf{q})$, then the modulus square of $A(\mathbf{q})$ will give us the intensities in the reciprocal space as $I(\mathbf{q})=|A(\mathbf{q})|^2$ as in FIG.6(c). This can reproduce the image as in FIG.6(d) formed by electron densities of the object that we have already found in (15) after phasing [3][4] without any constraint in the real space. The length of the major axis of the image appeared approximately twice as long as the minor one.

\begin{figure}
\includegraphics[scale=0.5]{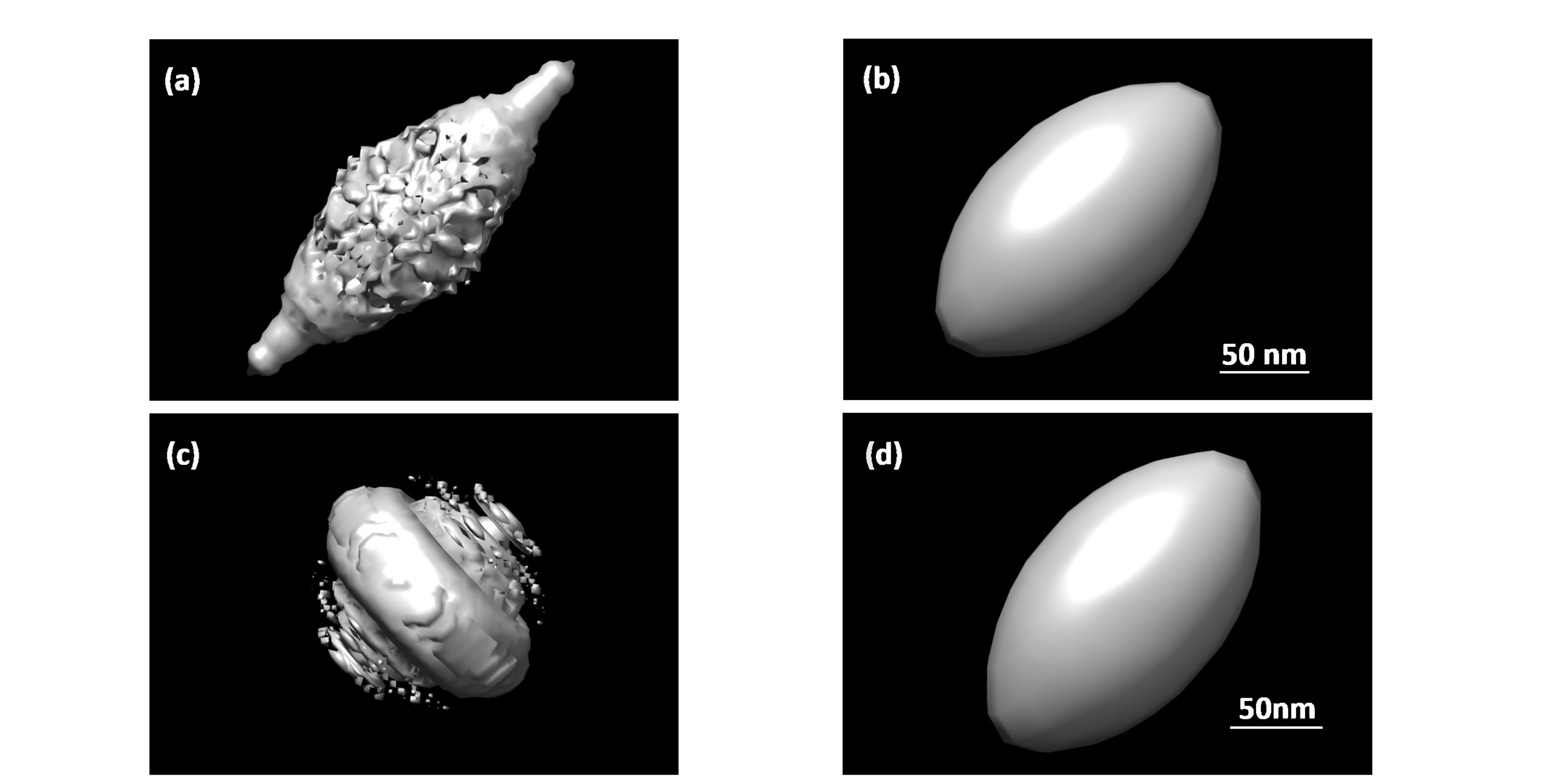}
\label{FIG.6}
\caption{(a)The diffraction volume of the Nanorice particle in the reciprocal space projected onto 449x449x449 cubic grids. (b) The electron densities in the real space projected onto 29x29x29 cubic grids. (c) The intensities formed by $A_{l,0}(q)$ that are derived from the electron densities $\rho_{l,0}(r)$ with (17), and projected onto 29x29x29 reciprocal space grids. (d) The 3D reconstruction image (electron densities) after phasing the intensities $I(\mathbf{q})=|A(\mathbf{q})|^2$ through (18) projected onto 29x29x29 real space cubic grids.}
\end{figure} 

The electron densities that we found from the diffraction patterns of the Nanorice particle ($Fe_2O_3$) can give us the average resolution of the image FIG.7(a) through the Fourier Shell Correlation (FSC) [15] defined by

\begin{equation}
FSC(q)=\frac{\sum\limits_{q_i\in q}A_1(q_i)A_2^{*}(q_i) }{\sqrt{\left[\sum\limits_{q_i\in q}|A_1(q_i)|^2 \right] \left[\sum\limits_{q_i\in q}|A_2(q_i)|^2\right]}}
\end{equation}

where $A_1$ and $A_2$ are the Fourier transforms of the electron densities of two randomly selected subsets of data. At this point, we have to consider $R_{split}(q)$ as in FIG.7(b) that would be a useful data quality indicator in x-ray diffraction, defined by [16]

\begin{equation}
R_{split}(q)=\frac{\sqrt{2}\sum\limits_{q_i \in q}|I_1(q_i)-I_2(q_i)|}{\sum\limits_{q_i \in q}|I_1(q_i)+I_2(q_2)|}
\end{equation}

The plot FIG.7 shows FSC(q) and Rsplit(q).

\begin{figure}
\includegraphics[scale=0.525]{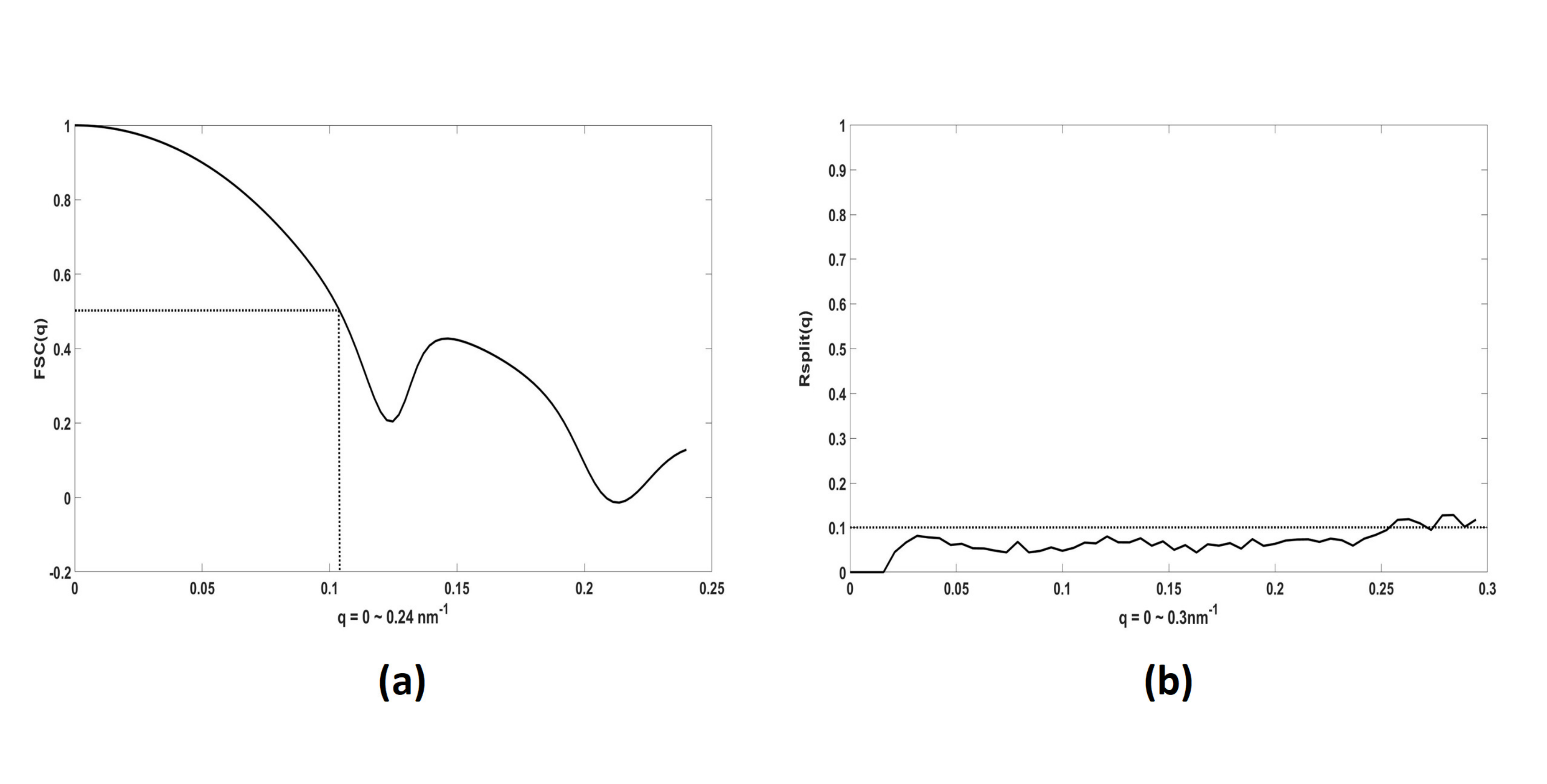}
\label{FIG.7}
\caption{(a) Plot of FSC(q) vs. q. (b) Rsplit(q) vs. q. q is ranged from 0 to 0.3$nm^{-1}$ where is approximately the edge of each diffraction pattern.}
\end{figure} 

Through the calculation of FSC(q) the average resolution of this electron density shows approximately 60nm while the best one is about 22nm at the edge of the diffraction patterns. Rsplit(q) as a data quality indicator shows almost all values are under 0.1 except at the edge of each DP where the intensities become blurred. This means these experimental data can be trusted enough to access the procedure of 3D imaging calculations. At this stage we point out that a small number of good quality data, as we used here with 98 DPs, is more meaningful than a large number of bad ones to access the 3D conformational imaging process of the reconstruction of nanoscale objects using the angular correlation method. \\

We acknowledge support for this work from a National Science Foundation Science and Technology Center (NSF Grant No. 1231306) and the UWM High Performance Computing Center (HPC) for the use of AVI and MORTIMER. Portion of this research were carried out at the Linac Coherent Light Source (LCLS) at the SLAC National Accelerator Laboratory. LCLS is an Office of Science User Facility operated for the US Department of Energy Office of Science by Stanford University. Use of the Linac Coherent Light Source (LCLS), SLAC National Accelerator Laboratory, is supported by the U.S. Department of Energy, Office of Science, Office of Basic Energy Science under Contract No. DE-AC02-76SF00515.


\end{document}